\documentstyle[prl,preprint,aps,epsf]{revtex}
\begin{document}
\title{Compromise of Localized Graviton with a Small Cosmological Constant
in Randall-Sundrum Scenario} 
%\preprint{MCTP-01-19}
\author{
D. K. Park\raisebox{0.8ex}{1,2}\footnote[1]
{Email:dkpark@hep.kyungnam.ac.kr
} and 
S. Tamaryan\raisebox{0.8ex}{3,4}\footnote[2]{Email:sayat@physik.uni-kl.de,
sayat@moon.yerphi.am}}
\address{$^1$ Department of Physics, Kyungnam University, Masan, 631-701,
	      Korea   \\
	 $^2$ Michigan Center for Theoretical Physics \\
	 Randall Laboratory, Department of Physics, University of Michigan \\
          Ann Arbor, MI 48109-1120, USA \\
	  $^3$ Theory Department, Yerevan Physics Institute, Yerevan--36,
	  375036, Armenia \\
	  $4$ Department of Physics, University of Kaiserslautern,
	  D-67653 Kaiserslautern, Germany}

\maketitle

%\date{\today}
\maketitle
\begin{abstract}
A  new mechanism which leads to a linearized massless graviton 
localized on the
brane is found in the $AdS$/CFT setting, {\it i.e.} in  a  single copy of 
$AdS_5$ spacetime with a singular brane on the boundary, within the
Randall-Sundrum brane-world scenario. With an  help  of a 
recent development in   path-integral techniques, a  one-parameter family
of propagators for linearized gravity is obtained  analytically, in which  
a  parameter $\xi$ reflects
various kinds  of  boundary conditions that  arise
as a result of the  half-line constraint. 
In the case of a Dirichlet boundary condition ($\xi = 0$) 
the graviton localized on the brane can be massless 
{\it via}
coupling constant renormalization. Our result supports a conjecture that 
the usual Randall-Sundrum scenario is a regularized version of a certain
underlying theory.
\end{abstract}

%---------------------------------------------------------------------------
%\newpage
\vspace{2cm}
The most remarkable feature of the
Randall-Sundrum(RS) brane-world scenario is
that it leads to a massless graviton localized on
 the 3-brane at the linearized
fluctuation level \cite{rs99}. In fact, this striking feature seems to 
furnish a motivation for   the recent application of this scenario to 
various branches of physics such as 
cosmology \cite{bine00,csa99-1,cli99,kanti99,csa99-2},
 the cosmological constant
hierarchy \cite{kim00,kim01,alex01}, and blackhole 
physics \cite{emp99,gid00,emp00}. The fact that RS spacetime is composed of the
two copies of $AdS_5$ attached along the boundary($y=0$) also provides 
another motivation for  the recent 
activity on the relation of this scenario to 
$AdS$/CFT
\cite{gid00,verl99,lykk99,gub99,duff00,anc00,deg00,pere01}.

When solving a linearized fluctuation equation, however, the authors of
Ref.\cite{alex01} chose  a Dirichlet boundary condition (BC) on the brane
to explain a small cosmological constant of the brane. In this BC the
3-brane acts effectively as a perfectly reflecting mirror,  and the 
cosmological constant 
becomes naturally very small  through   thermal radiation 
of vacuum energy from the brane
into the bulk. Then, it is very unclear why two different BCs are necessary
to explain two distinct phenomena. 
There should exist a single physical BC which explains these two different 
phenomena simultaneously.
In this context it is important to 
find a possible compromise of these two phenomena, which is 
a purpose of this letter.
As will be shown 
below, there exists a novel mechanism which leads to a massless physical 
graviton with  the Dirichlet BC 
{\it via}  coupling constant renormalization in the $AdS$/CFT
setting, {\it i.e.} that of a single $AdS_5$ spacetime with a singular brane
on the boundary. We argue here that the mixture procedure of Dirichlet
BC and the coupling constant renormalization is a most probable candidate
for the compromise. It also makes us conjecture that RS scenario is a
regularized version of a certain underlying theory.

Recently, it was  shown\cite{park01} that at nonzero temperature only 
half of the full spacetime
in the
RS scenario becomes Schwarzschild-$AdS_5$ due to the manifest $Z_2$-symmetry
breaking. 
Therefore, the choice of a single $AdS_5$ in the 
RS scenario also guarantees that
the close relationship of the RS scenario with $AdS$/CFT is maintained at
finite temperature.

We start with the gravitational fluctuation equation\cite{rs99} in the  RS
scenario, {\it i.e.} 
\begin{eqnarray}
\label{fluctua}
& &\hat{H}_{RS} \hat{\psi}(z) = \frac{m^2}{2} \hat{\psi}(z),  \\  \nonumber
& &\hat{H}_{RS} = -\frac{1}{2} \partial_z^2 + 
		\frac{15}{8(|z| + \frac{1}{k})^2} - \frac{3}{2} k \delta(z),
\end{eqnarray}
where $\hat{\psi}(z)$ is  related to  a linearized gravitational
field $h(\bar{x}, y)$ as follows:
\begin{eqnarray}
\label{relation1}
h(\bar{x}, y)&=&\psi(y) e^{ip\bar{x}},  \\   \nonumber
\hat{\psi}(z)&=&\psi(y) e^{\frac{k|y|}{2}}
\end{eqnarray}
where $z=\epsilon(y)(e^{k|y|} - 1) / k$, $p^2 = - m^2$, and $\bar{x}$ is 
the worldvolume coordinate. 
Since all components are the same, the Lorentz indices are suppressed in 
Eq.(\ref{relation1}).
When deriving the fluctuation equation, RS used the 
gauge choice
\begin{equation}
\label{gauge}
h_{55}=h_{\mu 5}=0, 
\hspace{1.0cm}
h_{\mu}^{\nu}{}_{,\nu} =0,  h_{\mu}^{\mu} = 0
\end{equation}
where $\mu, \nu = 0, 1, 2, 3$.
However, the choice of this gauge in the bulk generates in general a 
non-trivial bending structure of the 3-brane which is fully discussed in 
Refs.\cite{gid00,garr00,kaku01}. Since it does not change the main conclusion,
we will not explore the
subtlety of this gauge choice in detail here. 
What we want to do is to 
examine the properties of the Feynman
propagator explicitly for the general Hamiltonian 
\begin{eqnarray}
\label{defhamil}
\hat{H}&=&\hat{H}_0 - v \delta(z),   \\   \nonumber
\hat{H}_0&=& -\frac{1}{2} \partial_z^2 + 
	       \frac{g}{(|z| + c)^2},
\end{eqnarray}
when $z$ is non-negative. Of course,
$\hat{H}$ coincides with $\hat{H}_{RS}$ when $g=15/8$, $c=1/k\equiv R$, and 
$v=3k/2$,  where $R$ is the radius of $AdS_5$. 

From the purely  mathematical point of view
 the Hamiltonian $\hat{H}$ is a 
singular operator due to its point interaction. While the proper treatment
of the one-dimensional $\delta$-function potential in the
 Schr\"{o}dinger picture
was found long ago \cite{kron34}, it is  not quite so long ago
that one    understood how to 
treat it within the path-integral formalism. Following 
Schulman's procedure\cite{gave86,schul86}, it is possible to express the fixed-energy amplitude
$\hat{G}[z_1,z_2:E]$ for $\hat{H}$ in terms of the fixed-energy amplitude
$\hat{G}_0[z_1,z_2:E]$ for $\hat{H}_0$ as follows\footnote{The definition of 
the fixed-energy amplitude $\hat{G}[x,y:E]$ in this letter is a  Laplace
transform  of the usual Euclidean Feynman propagator $G[x,y:t]$.}
\begin{equation}
\label{g0g}
\hat{G}[z_1, z_2: E] = \hat{G}_0[z_1, z_2: E] + 
\frac{\hat{G}_0[z_1, 0: E] \hat{G}_0[0, z_2: E]}
     {\frac{1}{v} - \hat{G}_0[0, 0, E]}.
\end{equation}
The remaining problem, therefore, is to compute a fixed-energy amplitude 
for the 
Hamiltonian $\hat{H}_0$. 

As mentioned before, we would like to use only  half of the full RS
spacetime for the computation of $\hat{G}_0[z_1, z_2: E]$.  
In this case the fixed-energy amplitude is in general dependent upon the 
BC at the boundary arising  due to the  half-line constraint,
$z \geq 0$. In this half-line $\hat{H}_0$ becomes simply
\begin{equation}
\label{hamil0}
\hat{H}_0 = -\frac{1}{2} \partial_x^2 + \frac{g}{x^2}
\end{equation}
where $x = z + c$. Thus  our half-line constraint $z \geq 0$ is changed into 
$x \geq c$. If $c = 0$, the Euclidean propagator $G_{>0}[a, b: t]$ and 
the corresponding fixed-energy amplitude $\hat{G}_{>0}[a, b: E]$ 
for Hamiltonian (\ref{hamil0}) are
well-known \cite{schul81}:
\begin{eqnarray}
\label{wcis0}
G_{>0}[a, b: t]&=&\frac{\sqrt{a b}}{t} e^{-\frac{a^2 + b^2}{2 t}}
		  I_{\gamma}\left(\frac{ab}{t}\right),  \\  \nonumber
\hat{G}_{>0}[a, b: E]&=&2 \sqrt{a b} I_{\gamma} \left( \sqrt{\frac{E}{2}}
					((a + b) - |a - b|) \right)
                         K_{\gamma} \left( \sqrt{\frac{E}{2}}
					 ((a + b) + |a - b|) \right),
\end{eqnarray}
where $I_{\gamma}(z)$ and $K_{\gamma}(z)$ are the  usual modified Bessel
functions, and $\gamma = \sqrt{1 + 8g} / 2$. 

The difficulty of the computation of the fixed-energy amplitude for 
$\hat{H}_0$ is mainly due to the fact that the constraint
is not half-line in terms of $x$ but $x > c$, {\it i.e.}
$\hat{G}_0[a, b: E] = \hat{G}_{>c}[a,b:E]$. It may  be extremely difficult
to compute a path-integral directly with our asymmetric constraint. 
In this letter, instead of this direct approach,  we adopt the  following 
technique to solve the problem. First, 
we impose the  usual half-line constraint $x > 0$. 
Then, we introduce an infinite energy barrier at $x = c$ in $\hat{H}_0$ to
forbid a penetration into the region
 $0 < x < c$. The infinite energy barrier
can be consistently introduced  within the path-integral formalism using
$\delta$- and $\delta^{\prime}$-functions by assuming  an
infinitely large  coupling constant
\cite{grosch93,grosch95,grosch98}.
For the Dirichlet
and Neumann BC cases the fixed-energy amplitudes 
$\hat{G}_0^D[a,b:E]$ and $\hat{G}_0^N[a,b:E]$ for $\hat{H}_0$ 
with the infinite barrier are obtained
from $\hat{G}_{>0}[a, b: E]$ as follows:
\begin{eqnarray}
\label{gdgn}
\hat{G}_0^D[a, b: E]&=& \hat{G}_{>0}[a, b: E] - 
\frac{\hat{G}_{>0}[a, c: E] \hat{G}_{>0}[c, b: E]}
     {\hat{G}_{>0}[c^+, c: E]},                     \\   \nonumber
\hat{G}_0^N[a, b: E]&=& \hat{G}_{>0}[a, b: E] -
\frac{\hat{G}_{>0, b}[a, c: E] \hat{G}_{>0, a}[c, b: E]}
     {\hat{G}_{>0, ab}[c^+, c: E]}
\end{eqnarray}
where we used a point-splitting method to avoid an infinity arising  in 
$\hat{G}_{>0}[a, b: E]$ and $\hat{G}_{>0, ab}[a, b: E]$ at $a = b$.

The quantities
$\hat{G}_0^D[a, b: E]$ and $\hat{G}_0^N[a, b: E]$ are straightforwardly
computed using Eq.(\ref{gdgn}). The  explicit expressions are
\begin{eqnarray}
\label{explicit}
\hat{G}_0^D[a, b: E]&=&\hat{G}_{>0}[a, b: E] - 
2 \sqrt{ab} \frac{I_{\gamma}(\sqrt{2E} c)}{K_{\gamma}(\sqrt{2E} c)}
K_{\gamma}(\sqrt{2E} a) K_{\gamma}(\sqrt{2E} b),
						 \\   \nonumber
\hat{G}_0^N[a, b: E]&=&\hat{G}_{>0}[a, b: E] + 
2 \sqrt{ab} \frac{f_I(E)}{f_K(E)} 
K_{\gamma}(\sqrt{2E} a) K_{\gamma}(\sqrt{2E} b)
\end{eqnarray}
where
\begin{eqnarray}
\label{fkfi}
f_K(E)&=&\frac{\gamma - \frac{1}{2}}{\sqrt{2 E} c} K_{\gamma}(\sqrt{2E} c)
	 + K_{\gamma - 1}(\sqrt{2E} c),
							       \\  \nonumber
f_I(E)&=&I_{\gamma - 1}(\sqrt{2E} c) -
\frac{\gamma - \frac{1}{2}}{\sqrt{2 E} c} I_{\gamma}(\sqrt{2E} c).
\end{eqnarray}
It is simple to show that $\hat{G}_0^D[a,b:E]$ and $\hat{G}_0^N[a,b:E]$
satisfy the usual Dirichlet and Neumann BCs at $x = c$.

One may impose a mixing of Dirichlet and Neumann  BCs at $x = c$. 
In this case 
the fixed-energy amplitude
$\hat{G}_0[a, b: E]$ for $\hat{H}_0$ becomes a one parameter family 
of propagators\footnote{The boundary condition for the one-dimensional 
singular operator involves in general four real self-adjoint 
parameters \cite{alb94,park96}. In this letter, however, we do not 
explore this purely  mathematically  oriented approach.}
\begin{equation}
\label{genbc}
\hat{G}_0[a, b: E] = \xi \hat{G}_0^N[a, b: E] + (1 - \xi) \hat{G}_0^D[a, b: E]
\end{equation}
where $0 \leq \xi \leq 1$. Of course, the cases
 $\xi = 0$ and $\xi = 1$  correspond
to 
pure Dirichlet and pure Neumann BC cases. Another interesting case  
is the value $\xi = 1/2$, in which  the contributions of Neumann and Dirichlet
have equal weighting factors. Since $\hat{G}_0[a, b: E]$ is expressed
 in terms 
of eigenvalues $E_n$ and eigenfunctions  $\phi_n$ of $\hat{H}_0$ as follows
\begin{equation}
\label{rg0phi}
\hat{G}_0[a, b: E] = \sum_n \frac{\phi_n(a) \phi_n^{\ast}(b)}{E - E_n},
\end{equation}
the $\xi = 1/2$ case should correspond to the gravitational propagator
without any constraint in $x$. As will be shown below, this case exactly
reproduces the original RS  result.

Inserting (\ref{genbc}) into (\ref{g0g}) one can finally obtain the 
$\xi$-dependent propagator for $\hat{H}$ whose explicit form is 
\begin{eqnarray}
\label{xi-dep}
\hat{G}[a, b: E]&=& 2 \sqrt{a b} \Bigg[ I_{\gamma}(\sqrt{2E} min(a, b))
				       K_{\gamma}(\sqrt{2E} max(a, b))
							       \\  \nonumber
                  &+& \frac{K_{\gamma}(\sqrt{2E} a) K_{\gamma}(\sqrt{2E} b)}
			  {f_K(E)}
                     \bigg[ \xi \left( f_I(E) + \frac{1}{cE}
				     [\frac{f_K(E)}{\xi v} - \sqrt{\frac{2}{E}}
				      K_{\gamma}(\sqrt{2E}c)]^{-1}\right) 
							       \\  \nonumber
& & \hspace{3.0cm}
			   - (1 - \xi) \frac{I_{\gamma}(\sqrt{2E} c)}
					     {K_{\gamma}(\sqrt{2E} c)}
					     f_K(E)    
							       \bigg]
								 \Bigg].
\end{eqnarray}

We now  consider special cases of Eq.(\ref{xi-dep}). As expected, taking 
$\xi = 1/2$ with $g = v = 0$ makes $\hat{G}[a, b: E]$  the exact
free-particle amplitude. If one takes the
RS limit $g = 15/8$, $c = 1/k=R$, $v = 3k/2$ and $E=m^2/2$ at the same 
$\xi$ value, 
it is  possible to show that Eq.(\ref{xi-dep}) yields
\begin{equation}
\label{rspropa1}
\hat{G}^{RS}[a=R, b: m] = \frac{1}{m} \sqrt{\frac{b}{R}}
\frac{K_2(mb)}{K_1(mR)}.
\end{equation}
If we takes $b = R$, the amplitude becomes simply 
\begin{equation}
\label{rspropa2}
\hat{G}^{RS}[R, R: m] = R (\Delta_0 + \Delta_{KK}),
\end{equation}
where $\Delta_0$ and $\Delta_{KK}$ represent zero-mass localized gravity and
higher Kaluza-Klein excitation
\begin{eqnarray}
\label{D0DKK}
\Delta_0&=& \frac{2}{m^2 R^2},   \\  \nonumber
\Delta_{KK}&=& \frac{1}{mR} \frac{K_0(mR)}{K_1(mR)},
\end{eqnarray}
respectively.
 In this case,  when the separation between masses on the  brane is very
large, Newton's law  becomes
\begin{equation}
\label{rsnewton}
V_{RS} \sim \frac{1}{r} \left[ 1 + \left(\frac{R}{r}\right)^2\right]
\end{equation}
which agrees  with the RS result\cite{rs99}.

The first term in Eq.(\ref{rsnewton}) is a usual Newton potential contributed 
from the zero mode $\Delta_0$. The second term represents the correction to the
potential and is generated from the Kaluza-Klein excitation $\Delta_{KK}$. It is 
worthwhile noting that the correction to the potential is also computed
in Ref.\cite{garr00} using somewhat different method and the final result is 
different from Eq.(\ref{rsnewton}):
\begin{equation}
\label{revise1}
V_{RS} \sim \frac{1}{r} \left[ 1 + \frac{2}{3}\left(\frac{R}{r}\right)^2\right].
\end{equation}
The $2/3$ factor in Eq.(\ref{revise1}) is derived by considering the source term
arising from the bending structure of the $3$-brane. 
Thus, the factor difference in potential is due to our ignorance of the 
bending effect.
It is interesting to examine 
how to involve the bending effect within the path-integral formalism.

If we choose $\xi = 1$ with RS limit, 
$\hat{G}[a, b: E]$ of
Eq.(\ref{xi-dep}) 
 reduces to
\begin{equation}
\label{neupropa}
\hat{G}^N[R, R: E] = 2 R \frac{\Delta}{1 - \frac{3}{2} \Delta}
\end{equation}
where $\Delta = \Delta_0 + \Delta_{KK}$. Numerical calculation shows that 
there exists a massive graviton bound on the brane in this case whose mass is 
\begin{equation}
\label{neumass}
m_N \approx 2.48 R^{-1}.
\end{equation}
It is well-known that the potential due to the exchange of a massive particle
is exponentially suppressed at long distance.
This result
is reasonable because the massive 
particle in general cannot propagate a long distance freely.

Finally, we  consider the case $\xi = 0$.
In this case the result  (\ref{xi-dep})    of the
usual Schulman  
procedure  does not yield an any modification 
due to the Dirichlet BC if the coupling constant $v$ is finite.
As shown in \cite{jack91,park95}, however,  we can obtain a non-trivial 
modification of the fixed-energy
amplitude in this case {\it via}  coupling constant 
renormalization if $v$ is an infinite bare quantity. 
In this letter we will follow this procedure by treating $v$ as an unphysical
infinite quantity. This means we abandon the RS limit $v=3k/2$ at $\xi=0$
case. As will be shown shortly, this procedure also generates a massless 
gravity localized on the brane when the renormalized coupling constant 
becomes a particular value.

To show this more explicitly
we introduce a positive infinitesimal parameter $\epsilon$ for the
regularization and rewrite Eq.(\ref{g0g}) in the form:
\begin{equation}
\label{g0g-1}
\hat{G}^D[a, b: E] = \hat{G}_0^D[a, b: E] + 
\lim_{\epsilon \rightarrow 0^+}
\frac{\hat{G}_0^D[a, c + \epsilon: E] \hat{G}_0^D[c + \epsilon, b: E]}
     {\frac{1}{v} - \hat{G}_0^D[c + \epsilon, c + \epsilon: E]}.
\end{equation}
Using the expansions
\begin{eqnarray}
\label{dexpand}
\hat{G}_0^D[a, c + \epsilon: E]&=&2 \sqrt{\frac{a}{c}} 
\frac{K_{\gamma}(\sqrt{2E} a)}{K_{\gamma}(\sqrt{2E} c)} \epsilon
+ O(\epsilon^2),    
				       \\   \nonumber
\hat{G}_0^D[c + \epsilon, b: E]&=& 2 \sqrt{\frac{b}{c}}
\frac{K_{\gamma}(\sqrt{2E} b)}{K_{\gamma}(\sqrt{2E} c)} \epsilon
+ O(\epsilon^2),    
				       \\   \nonumber
\hat{G}_0^D[c + \epsilon, c + \epsilon: E]&=& 2 \epsilon + 
\frac{2 \epsilon^2}{c} \Omega(\sqrt{2E} c, \gamma) + O(\epsilon^3),
\end{eqnarray}
where
\begin{equation}
\label{defomega}
\Omega(z, \nu) = 1 + z \frac{K_{\nu}^{\prime}(z)}{K_{\nu}(z)}
		+ \frac{z^2}{2} (I_{\nu}^{\prime \prime}(z) K_{\nu}(z)
				 - I_{\nu}(z) K_{\nu}^{\prime \prime}(z)),
\end{equation}
it is straightforward to derive a non-trivial fixed-energy amplitude
\begin{eqnarray}
\label{dipropa}
\hat{G}^D[a, b: E]&=& 2 \sqrt{a b}
\Bigg[I_{\gamma}[\sqrt{2E} min(a, b)] K_{\gamma}[\sqrt{2E} max(a, b)]
							      \\  \nonumber
     &-& \frac{K_{\gamma}(\sqrt{2E} a) K_{\gamma}(\sqrt{2E} b)}
	     {K_{\gamma}^2(\sqrt{2E} c)}
	     \left[ I_{\gamma}(\sqrt{2E} c) K_{\gamma}(\sqrt{2E} c)
		   + \frac{1}{2[\Omega(\sqrt{2E} c, \gamma) - v^{ren} c]}
							\right] \Bigg]
\end{eqnarray}
where the renormalized coupling constant $v^{ren}$ is defined in terms of the 
bare coupling constant
as follows:
\begin{equation}
\label{reandba}
v^{ren} = \frac{1}{2 \epsilon^2} \left( \frac{1}{v} - 2 \epsilon \right).
\end{equation}
One can easily show  that $v^{ren}$ has the  same dimension as the  bare 
coupling constant $v$. 
Following the philosophy of renormalization we regard $v^{ren}$ as a finite
quantity. Taking the remaining RS limit $g = 15/8$, $c = 1/k = R$, 
and $E = m^2 / 2$,
one can show that the fixed-energy amplitude in this case is 
\begin{equation}
\label{dddd}
\hat{G}^D[R, b: E] = \sqrt{R b} \frac{K_2(mb)}{K_2(mR)}
\frac{1}{v^{ren} R - \Omega(mR, 2)}.
\end{equation}
Using
\begin{equation}
\label{omega2}
\Omega(mR, 2) = -\frac{3}{2} - m R \frac{K_1(mR)}{K_2(mR)}
\end{equation}
it is possible to show that the corresponding gravitational potential 
at long range is 
%\begin{equation}
%\label{dgra1}
%V_D \sim \frac{2 R}{(R v^{ren} + \frac{3}{2}) r^3}
%\left[ 1 - \frac{6}{R v^{ren} + \frac{3}{2}} \left(\frac{R}{r}\right)^2 \right]
%\end{equation}
%when $R v^{ren} + 3/2 \neq 0$ and 
\begin{equation}
\label{dgra2}
V_D \sim \frac{1}{r} \left[ 1 +  \left(\frac{R}{r}\right)^2
						  \right] = V_{RS}
\end{equation}
when $R v^{ren} + 3/2 = 0$. 
Hence, we obtain a massless graviton localized on the brane when 
$v^{ren} = -3 / (2 R)$. Eq.(\ref{dgra2}) is a surprising result.
 Although
we obtained a massless graviton through completely different BC and completely
different procedure, its gravitational potential on the 3-brane is exactly the 
same as that of the  original RS  result.
 This exact coincidence strongly supports
the conjecture that the  Dirichlet BC for Hamiltonian $\hat{H}_0$
 is a genuine physical BC in the linearized 
gravity theory of RS scenario.
The requirement of the coupling constant renormalization supports 
another conjecture that RS
scenario is a regularized version of a certain underlying theory. It would 
be interesting to find and examine the underlying theory which might be
our future work.

At $v^{ren} = -3 / (2R)$ the graviton propagator (\ref{dipropa})  
reduces to the following simple form in the  RS limit
\begin{equation}
\label{final1}
\hat{G}^D[a, b: m] = \hat{G}_0^D[a, b: m] + \sqrt{a b}
\frac{K_2(m a) K_2(m b)}{K_2^2(m R)} 
\left(\Delta_0 + \Delta_{KK} \right).
\end{equation}
The first term in Eq.(\ref{final1}) is responsible for the small cosmological
constant through thermal radiation of vacuum energy from the brane
into the bulk due to its Dirichlet 
nature \cite{alex01}. The second term is responsible for the massless 
graviton localized on the brane. Of course, because of the second term
the $3$-brane cannot act as a perfectly reflecting mirror in the bulk. 
This  may be a physical reason why the cosmological constant of our
universe is nonzero. Therefore, it might be also interesting to estimate 
the value
of the cosmological constant within the present scenario and compare it with 
real experimental data $(0.01 eV)^4$.

\end{document}